\documentclass[journal=jacsat,manuscript=article]{achemso}

\usepackage{chemformula} % Formula subscripts using \ch{}
\usepackage[T1]{fontenc} % Use modern font encodings
\author{Jian Tang}
\affiliation{Key Laboratory of Low-Dimensional Quantum Structures and Quantum Control of Ministry of Education, Department of Physics and Synergetic Innovation Center for Quantum  Effects and Applications, Hunan Normal University, Changsha 410081, China}

\author{Yun-Lan Zuo}
\affiliation{Key Laboratory of Low-Dimensional Quantum Structures and Quantum Control of Ministry of Education, Department of Physics and Synergetic Innovation Center for Quantum Effects and Applications, Hunan Normal University, Changsha 410081, China}

\author{Xun-Wei Xu}
\affiliation{Key Laboratory of Low-Dimensional Quantum Structures and Quantum Control of Ministry of Education, Department of Physics and Synergetic Innovation Center for Quantum Effects and Applications, Hunan Normal University, Changsha 410081, China}

\author{Ran Huang}
\email{ran.huang@riken.jp}
\affiliation{Theoretical Quantum Physics Laboratory, Cluster for Pioneering Research, RIKEN, Wako-shi, Saitama 351-0198, Japan}
\alsoaffiliation{Quantum Information Physics Theory Research Team, Quantum Computing Center, RIKEN, Wako-shi, Saitama 351-0198, Japan}

\author{Adam Miranowicz}
\affiliation{Theoretical Quantum Physics Laboratory, Cluster for Pioneering Research, RIKEN, Wako-shi, Saitama 351-0198, Japan}
\alsoaffiliation{Institute of Spintronics and Quantum Information, Faculty of Physics, Adam Mickiewicz University, 61-614 Pozna{\'{n}}, Poland}

\author{Franco Nori}
\affiliation{Theoretical Quantum Physics Laboratory, Cluster for Pioneering Research, RIKEN, Wako-shi, Saitama 351-0198, Japan}
\alsoaffiliation{Quantum Information Physics Theory Research Team, Quantum Computing Center, RIKEN, Wako-shi, Saitama 351-0198, Japan}
\alsoaffiliation{Physics Department, The University of Michigan, Ann Arbor, Michigan 48109-1040, USA}

\author{Hui Jing}
\email{jinghui73@foxmail.com}
\affiliation{Key Laboratory of Low-Dimensional Quantum Structures and Quantum Control of Ministry of Education, Department of Physics and Synergetic Innovation Center for Quantum Effects and Applications, Hunan Normal University, Changsha 410081, China}

\title[An \textsf{achemso} demo]
  {Achieving Robust Single-Photon Blockade with a Single Nanotip}


\begin{document}

%\begin{tocentry}

%Some journals require a graphical entry for the Table of Contents.
%This should be laid out ``print ready'' so that the sizing of the

%Inside the \texttt{tocentry} environment, the font used is Helvetica
%8\,pt, as required by \emph{Journal of the American Chemical

%The surrounding frame is 9\,cm by 3.5\,cm, which is the maximum
%permitted for  \emph{Journal of the American Chemical Society}
%graphical table of content entries. The box will not resize if the
%content is too big: instead it will overflow the edge of the box.

%This box and the associated title will always be printed on a
%separate page at the end of the document.

%\end{tocentry}

\begin{abstract}
Backscattering losses, due to intrinsic imperfections or external perturbations that are unavoidable in optical resonators, can severely affect the performance of practical photonic devices. In particular, for quantum single-photon devices, robust quantum correlations against backscattering losses, which are highly desirable for diverse applications, have remained largely unexplored. Here, we show that single-photon blockade against backscattering loss, an important purely quantum effect, can be achieved by introducing a nanotip near a Kerr nonlinear resonator with intrinsic defects. We find that the quantum correlation of single photons can approach that of a lossless cavity even in the presence of strong backscattering losses. Moreover, the behavior of such quantum correlation is distinct from that of the classical mean-photon number with different strengths of the nonlinearity, due to the interplay of the resonator nonlinearity and the tip-induced optical coupling. Our work sheds new light on protecting and engineering fragile quantum devices against imperfections, for applications in robust single-photon sources and backscattering-immune quantum devices.\end{abstract}

\section{Introduction}

Optical whispering-gallery-mode (WGM) microresonators, which confine a light wave in a circular path within a microscale volume, leading to intensely enhanced light-matter interactions, are significant for both fundamental studies of nonlinear optics~\cite{Lin2017Nonlinear,Strekalov2016Nonlinear} and applications in ultra-sensitive sensing~\cite{Tang2023Single,Mao2023whispering,Yu2021Whispering,Foreman2015Whispering}, or quantum communications~\cite{Furst2011Quantum,lu2019chip,Brooks2021Integrated}. In a real WGM cavity, imperfections---like intrinsic material defects, density variation, or surface roughness---can cause backscattered light in the counter-propagating direction, leading to an extra optical loss and mode coupling. Such backscattering has been used to realize counter-propagating solitons~\cite{Yang2017Counter}, chiral lasing~\cite{Peng2016Chiral} and absorption~\cite{Ren2023Backscattering}, as well as slow light and its localization~\cite{Lu2022High}. However, backscattering limits the application performance in classical and quantum devices, such as instability problems in frequency combs~\cite{suh2016microresonator,griffith2015silicon}, backscattering-induced noise, and lock-in effect in optical gyroscopes~\cite{Cutler1980Forward,Lai2020Earth,liang2017resonant}, as well as decrease of secure key rates in quantum key distribution~\cite{Patel2012Coexistence,Subacius2004Backscattering}.

Backscattering suppression was experimentally studied by introducing reflectors or scatterers, ranging from macroscale mirrors~\cite{Ang2017Fundamental,jaffe2022understanding} to Mie~\cite{lee2023chiral} and Rayleigh~\cite{svela2020coherent} scatterers. Also, Brillouin scattering~\cite{kim2019dynamic,orsel2023electrically}, active feedback control~\cite{krenz2007controlling}, and synthetic gauge fields~\cite{Chen2021Synthetic} were used to suppress backscattering. These remarkable achievements provide powerful tools to optimize optical devices~\cite{orsel2023electrically} and explore nonreciprocal optics~\cite{Chen2021Synthetic} or non-Hermitian physics~\cite{Jiang2024Coherent}. Yet previous efforts have been devoted to propagation against backscattering loss of {\it{many photons or classical light }}~\cite{Ang2017Fundamental,jaffe2022understanding,lee2023chiral,svela2020coherent,krenz2007controlling,Chen2021Synthetic,kim2019dynamic,orsel2023electrically,Jiang2024Coherent,Jiao2020Nonreciprocal,liu2023phase}, it is essential to study robust nonclassical single-photon effects in spite of intrinsic defects, which are expected to play a key role in realistic single-photon devices and quantum technologies.

Here we propose how to realize {\it{quantum correlations}} against backscattering loss via single-photon blockade (SPB) effect in a real WGM cavity with intrinsic imperfections. We note that SPB~\cite{Tian1992Quantum,Leo1994Possibility,Miranowicz1996Quantum,Imamo1997Strongly}, indicating blockade of the subsequent photons by absorbing the first one, goes beyond classical optics and laser physics into a purely quantum regime. In view of its important roles in generating nonclassical correlations and building single-photon devices, SPB has been demonstrated experimentally in various systems ranging from cavities with atoms~\cite{Birnbaum2005,Barak2008,Hamsen2018Strong}, quantum dots~\cite{hennessy2007quantum,Faraon2008,reinhard2012strongly,muller2015coherent,Snijders2018Observation}, or superconducting qubits~\cite{Lang2011Observation,Hoffman2011Dispersive,Vaneph2018Observation,Rolland2019Antibunched,Collodo2019Observation}, to cavity-free atoms~\cite{peyronel2012quantum,Prasad2020Correlating} or Bose-Hubbard chains~\cite{Fedorov2021Photon}. In addition, multi-photon blockade~\cite{hamsen2017two,Chakram2022Multimode,Chakram2021Seamless} has recently been observed, opening the way to create few-photon devices for quantum networks.

In this letter, we find that SPB against backscattering loss occurs in a nonlinear WGM cavity by tuning the tip position. The second-order quantum correlation is generated with a SPB against backscattering loss efficiency of up to $99.7\%$, and is robust with different backscattering strengths. Moreover, we find different types of behavior of quantum correlations and the classical mean photon number in the backscattering suppression process. Our findings do not merely rely on the scatterer-induced destructive interference~\cite{Ang2017Fundamental,jaffe2022understanding,lee2023chiral,svela2020coherent}, but on the {\it{interplay}} of the resonator nonlinearity and the tip-induced optical coupling. Instead of analyzing light amplitudes~\cite{Ang2017Fundamental,jaffe2022understanding,lee2023chiral,svela2020coherent,krenz2007controlling,Chen2021Synthetic,kim2019dynamic,orsel2023electrically,Jiang2024Coherent}, we focus on quantum correlations and the transitions between quantum states, which hold the potential for implementations in quantum information technologies. Our findings drive the field of backscattering suppression into the quantum regime, hence making it possible to realize a variety of quantum backscattering-immune effects, such as multi-photon blockade against backscattering loss~\cite{hamsen2017two,Chakram2022Multimode,Chakram2021Seamless} or one-way single-photon transmission~\cite{Dong2021All,Zou2023Unidirectional}, for applications in robust quantum devices and the protection of fragile quantum resources.

%%%%%%%%%%%%%%%%%%%%%%%%%%%%%%%%%%%%%%%%%%%%%%%%%%
\begin{figure}[t!]
    \includegraphics[width=1\textwidth]{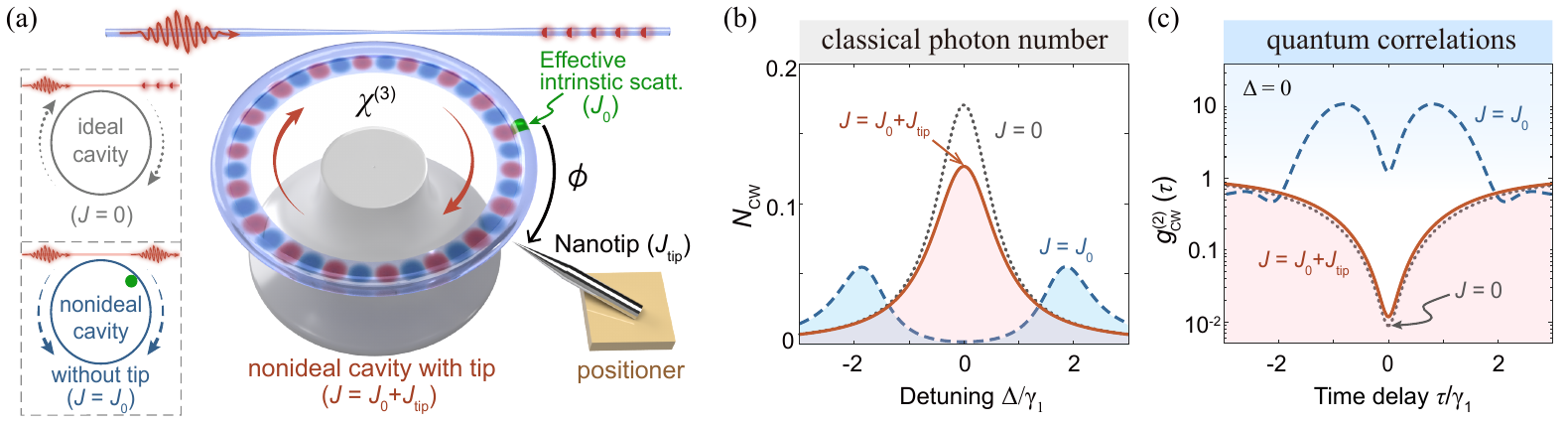}
    \caption{SPB against backscattering loss in a Kerr WGM cavity with an additional tip. (a) SPB occurs in an ideal nonlinear cavity (grey, $J=0$), and is annihilated by the backscattering in a nonideal cavity (blue, $J=J_0$). SPB reoccurs by tuning the relative distance $\phi$ between the intrinsic scatterer ($J_{0}$) and the tip ($J_{\textrm{tip}}$). (b) These effects are confirmed via the quantum correlation $g_{\textrm{cw}}^{(2)}(\tau)$. (c) Mean photon number $\textit{N}_{\textrm{cw}}$ versus $\Delta/\gamma_1$. Here, $J_{0}=1.8\gamma_1\sim0.4\ \mathrm{MHz}$, $\chi/\gamma\sim5.3$, $\phi=0.27\ \mu\mathrm{m}$. The other parameters are given in the main text.}
\label{Fig1}
\end{figure}
%%%%%%%%%%%%%%%%%%%%%%%%%%%%%%%%%%%%%%%%%%%%%%%%%%%
\section{Results and discussions}

\subsection{Model}
We consider an optical Kerr resonator with an additional nanotip [Fig.~\ref{Fig1}(a)]. For an ideal cavity driven from the left-hand side, only the clockwise (CW) mode is dominant. In a real cavity, intrinsic defects cause backscattering in the counterclockwise (CCW) direction, which can be approximated as an effective single scatterer~\cite{svela2020coherent,Matres2017Simple}, leading to the coupling between the CW and CCW modes (with strength $J_0$)~\cite{Jan2011Structure,Mohageg2007Coherent,li2016backscattering}. The intrinsic backscattering strength is proportional to $J_0$~\cite{Mazzei2007Controlled,Ren2023Backscattering}. By introducing a nanotip, the total optical coupling can be written as ($\hbar$ = 1)~\cite{svela2020coherent}:
\begin{align}\label{Eq:J21}
\hat{H}_{j} &= J\hat{a}_{1}^{\dagger }\hat{a}_{2}+J^*\hat{a}_{2}^{\dagger }\hat{a}_{1},\quad J=J_{0}+J_{\textrm{tip}}, \nonumber \\
J_{\mathrm{tip}} &={a_t}\mathrm{exp}(-2\beta_tr)\mathrm{exp}(-i\Theta).
\end{align}
Here, $\hat{a}_{1}$ ($\hat{a}_{2}$) is the annihilation operator for the CW (CCW) mode, $J_{\mathrm{tip}}$ is the tip-induced coupling strength with amplitude $a_{t}$, decay coefficient $\beta _t$, and radial distance $r$. The relative phase of the effective intrinsic scatterer and the tip is $\Theta=2k_{\mathrm{opt}}\phi+\theta+\theta_tr$, where $\phi$ is the relative azimuthal distance, $k_{\mathrm{opt}}=2\pi n_0/\lambda$ is the optical wavenumber with refractive index $n_0$ and vacuum wavelength of light $\lambda$, $\theta$ is the initial phase, and $\theta_t$ is a radially dependent phase accounting for the tip shape~\cite{svela2020coherent}.

To study SPB against backscattering loss, we consider a generic nonlinearity, Kerr nonlinearity~\cite{Leo1994Possibility,Imamo1997Strongly,miranowicz2013Twophoton}, which was realized via light-atom couplings~\cite{Birnbaum2005,Schmidt1996Absorption}, superconducting circuits~\cite{Kirchmair2013}, and magnon devices~\cite{Wang2018Bistability}, as well as theoretically studied in optomechanical systems~\cite{rabl2011photon,Nunnenkamp2011Single,Liao2013PB,gong2009Effective}. The Kerr interactions are given by~\cite{marin2017microresonator,cao2017Experimental,hales2018Third,Heuck2020Controlled}
\begin{align}
\hat{H}_{k} = \underset{j=1,2}{\sum}\chi\hat{a}_{j}^{\dagger }\hat{a}_{j}^{\dagger }\hat{a}_{j}\hat{a}_{j}+2\chi \hat{a}_{1}^{\dagger }\hat{a}_{1}\hat{a}_{2}^{\dagger }\hat{a}_{2},
\end{align}
where $\chi=3\hbar\omega^2\chi^{(3)}/(4\varepsilon_0\varepsilon_r^2V_\mathrm{eff})$ is the Kerr parameter with nonlinear susceptibility $\chi^{(3)}$, vacuum (relative) permittivity $\varepsilon_0$ ($\varepsilon_r$), and mode volume $V_\mathrm{eff}$~\cite{marin2017microresonator,cao2017Experimental,Choi2017Self, Boyd2008}. Such Kerr interaction becomes $\hat{H}_{k}=\chi\hat{a}_{1}^{\dagger }\hat{a}_{1}^{\dagger}\hat{a}_{1}\hat{a}_{1}$ in an ideal cavity. In the frame rotating at the drive frequency $\omega_L$, the Hamiltonian of the system reads
\begin{eqnarray}
\hat{H}_{r}=\Delta(\hat{a}_{1}^{\dagger }\hat{a}_{1}+\hat{a}_{2}^{\dagger }\hat{a}_{2})+\hat{H}_{j}+\hat{H}_{k}+\xi(\hat{a}_{1}^{\dagger }+\hat{a}_{1}),
\end{eqnarray}
with $\Delta=\omega-\omega_L$, and $\omega=\omega_0+\vert J\vert$. Here, $\omega_0$ is the resonance frequency of the cavity, $\xi=[\gamma_{\textrm{ex}} P_{\mathrm{in}}/\left(\hbar\omega_{L}\right)]^{1/2}$ is the driving amplitude with power $P_{\mathrm{in}}$ and cavity-waveguide coupling rate $\gamma_{\textrm{ex}}$.

% As shown in Fig.~\ref{Fig1}(a), single photons can be generated via SPB in an ideal cavity without backscattered light (upper inset), while the SPB is disrupted in an nonideal cavity with the presence of intrinsic backscattering (lower inset). Intriguingly, such SPB can reoccur by introducing an additional tip near to the nonideal cavity, indicating a quantum backscattering-immune effect on the single-photon level.
\subsection{SPB against backscattering loss}
We study the classical mean photon number $N_\mathrm{cw}=\langle\hat{a}_1^\dagger\hat{a}_1\rangle$, and the second-order quantum correlation~\cite{Glauber1963Quantum}: $g^{(2)}_{\mathrm{cw}}(\tau)\equiv\textstyle \lim_{t \to \infty} [\langle\hat{a}_1^\dagger(t)\hat{a}_1^\dagger(t+\tau)\hat{a}_1(t+\tau)\hat{a}_1(t)\rangle/\langle\hat{a}_1^\dagger(t)\hat{a}_1(t)\rangle^2]$, which is usually measured by Hanbury Brown-Twiss interferometers~\cite{Birnbaum2005,Barak2008,Hamsen2018Strong,hennessy2007quantum,Faraon2008}. The condition $g^{(2)}_{\mathrm{cw}}(0) < g^{(2)}_{\mathrm{cw}}(\tau)$ characterizes photon antibunching, and $g^{(2)}_{\mathrm{cw}}(0)\ll1$ [or $g^{(2)}_{\mathrm{cw}}(0)\approx0$] indicating SPB with sub-Poissonian photon-number statistics~\cite{Imamo1997Strongly,Birnbaum2005,scully1997quantum,agarwal2012quantum}.

This $g^{(2)}_{\mathrm{cw}}(\tau)$ can be calculated by numerically solving the Lindblad master equation for the density operator $\hat{\rho}$ of this system~\cite{johansson2012qutip,johansson2013qutip2}:
\begin{align}\label{eq:ME}
\dot{\hat{\rho}}=-i[\hat{H}_{r},\hat{\rho}]+\underset{j=1,2}{\sum}\frac{\gamma}{2}(2\hat{a}_{j}\hat{\rho}\hat{a}_{j}^{\dagger}-\hat{a}_{j}^{\dagger}\hat{a}_{j}\hat{\rho}-\hat{\rho}\hat{a}_{j}^{\dagger}\hat{a}_{j}),
\end{align}
where $\gamma=\gamma_{1}+\gamma_{\textrm{tip}}$ is the total dissipation rate, $\gamma_{1}=\gamma_{0}+\gamma_{\textrm{ex}}$, and $\gamma_{0}=\omega_0/Q$ denotes the intrinsic losses of the cavity with the quality factor $Q$. The tip-induced loss is $\gamma_{\textrm{tip}}=a_{\gamma}\mathrm{exp}(-2\beta_{\gamma}r)$, with amplitude $a_{\gamma}$, and decay coefficient $\beta_{\gamma}$~\cite{svela2020coherent}. The experimentally accessible parameters of the nanotip are taken as~\cite{svela2020coherent}: $a_{t}=14.3\ \mathrm{MHz}$, $(2{\beta_t})^{-1}=99\ \mathrm{nm}$, $\theta_t=3\pi/2\ \mu\mathrm{m}^{-1}$, $\theta=-\pi/2$, $a_{\gamma}=2.43\ \mathrm{MHz}$, and $(2\beta_{\gamma})^{-1}=92\ \mathrm{nm}$. The other experimentally accessible parameters are~\cite{vahala2003Optical, spillane2005Ultrahigh,pavlov2017Soliton,huet2016Millisecond,hales2018Third,Heuck2020Controlled}: $V_\mathrm{eff}=150\ \mu\mathrm{m}^3$, $n_0=1.4$, $Q=10^{10}$, $\lambda=1550\ \mathrm{nm}$, $\chi^{(3)}/\varepsilon_r^2=1.8\times10^{-17}\ \mathrm{m}^2/\mathrm{V}^2$, and $P_\mathrm{in}=4\ \mathrm{fW}$. For the WGM cavities, $V_\mathrm{eff}$ is typically $10^2$--$10^4\ \mu\mathrm{m}^3$~\cite{vahala2003Optical,spillane2005Ultrahigh}, $Q\sim10^9$--$10^{12}$~\cite{pavlov2017Soliton,huet2016Millisecond}, and $J_0\sim0.5\ \mathrm{MHz}$--$0.1\ \mathrm{GHz}$~\cite{Jin2021Hertz, kim2019dynamic,DelHaye2013Laser,orsel2023electrically}. The Kerr coefficient for semiconductor materials with GaAs is $\chi^{(3)}/\varepsilon_r^2=2\times10^{-17}\ \mathrm{m}^2/\mathrm{V}^2$~\cite{hales2018Third,Heuck2020Controlled}, and materials with indium tin oxide reach $\chi^{(3)}/\varepsilon_r^2=2.12\times10^{-17}\ \mathrm{m}^2/\mathrm{V}^2$~\cite{Alam2016Large}. In addition, $\chi^{(3)}$ can be further enhanced to $2\times10^{-11}\ \mathrm{m}/\mathrm{V}^2$ with other materials~\cite{zielinska2017self,Choi2017Self}. The input power can be attenuated by passing through an electro-optic modulator, and reach to $6.3\ \mathrm{fW}$~\cite{Wang2020}. Since $P_\mathrm{in}\ll\gamma$, the thermal effect induced by high optical powers can be neglected~\cite{svela2022near}. Thermal effects can also be reduced by making a thermal isolation or changing the materials of the tip and bracket to, for instance, aluminium~\cite{HAGARTALEXANDER2010269}. Also, realistic mechanical instability or temperature drift can be eliminated by designing a chip-based resonator with a scatterer integrated on the chip~\cite{Karabchevsky2020}, which holds the potential for realizing backscattering-immune on-chip resonators based on microelectromechanical-systems (MEMS) techniques.

%%%%%%%%%%%%%%%%%%%%%%%%%%%%%%%%%%%%%%%%%%%%%%%%%%
\begin{figure}[t!]
    \includegraphics[width=0.83\textwidth]{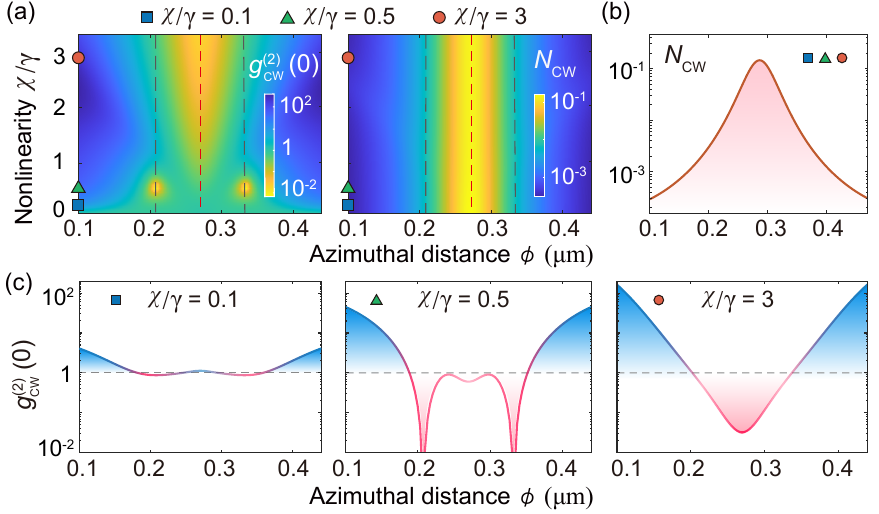}
    \caption{Different types of behavior of $g^{(2)}_\mathrm{cw}(0)$ and $\textit{N}_{\textrm{cw}}$ by varying $\phi$ and $\chi/\gamma$. (a) The $g_{\textrm{cw}}^{(2)}(0)$ relies on both of $\chi$ and $\phi$, while $\textit{N}_{\textrm{cw}}$ is independent of $\chi$. (b) At $\phi=0.27\ \mu\mathrm{m}$ [vertical red dashed line in (a)], $\textit{N}_{\textrm{cw}}$ always recovers, while SPB cannot revive for $\chi/\gamma<1$ (e.g., $\chi/\gamma=0.1$, blue squares), but can occur for $\chi/\gamma>1$ (e.g., $\chi/\gamma=3$, red circles). At $\phi=\{0.21,\ 0.33\}\ \mu\mathrm{m}$ [vertical gray dashed lines in (a)], SPB emerges for $\chi/\gamma=0.5$ (green triangles). The other parameters are the same as those in Fig.~\ref{Fig1}.}
\label{Fig2}
\end{figure}
%%%%%%%%%%%%%%%%%%%%%%%%%%%%%%%%%%%%%%%%%%%%%%%%%%%

Figure~\ref{Fig1}(b) shows SPB with $g^{(2)}_{\mathrm{cw}}(0)\sim0.009$ in an ideal Kerr cavity, since the input light fulfilling the single-photon resonance condition ($\Delta=0$) can only be resonant with the transition from the vacuum to the one-photon state but not with higher transitions~\cite{Imamo1997Strongly,Birnbaum2005}. However, intrinsic defects in a nonideal cavity cause backscattering and a coupling $J_0$ [the mode splitting in Fig.~\ref{Fig1}(c)], which provides an extra path for the resonance of higher-state transitions, leading to the breakdown of SPB.

In contrast, SPB recovers with a tip [Fig.~\ref{Fig1}(b)], which is also confirmed by higher-order correlations: $g^{(4)}_{\mathrm{cw}}(0)\sim4.8\times10^{-8}\ll g^{(3)}_{\mathrm{cw}}(0)\sim3.6\times10^{-5}\ll g^{(2)}_{\mathrm{cw}}(0)\sim0.012\ll 1$. Due to the tip-induced loss, $g^{(2)}_{\mathrm{cw}}(0)$ is slightly larger than that in the ideal cavity. Such quantum backscattering-immune effect is different from the classical one~\cite{Ang2017Fundamental,jaffe2022understanding,lee2023chiral,svela2020coherent,krenz2007controlling,Chen2021Synthetic,kim2019dynamic,orsel2023electrically,Jiang2024Coherent}.

%In contrast, SPB recovers with a tip [$g^{(2)}_{\mathrm{cw}}(0)\sim0.012$, Fig.~\ref{Fig1}(b)], which is also confirmed by higher-order correlations: $g^{(4)}_{\mathrm{cw}}(0)\ll g^{(3)}_{\mathrm{cw}}(0)\ll g^{(2)}_{\mathrm{cw}}(0)\ll 1$, \Revise{where $g^{(3)}_\mathrm{cw}(0)\sim3.6\times10^{-5}$, and $g^{(4)}_\mathrm{cw}(0)\sim4.8\times10^{-8}$}. Due to the tip-induced loss, $g^{(2)}_{\mathrm{cw}}(0)$ is slightly larger than that in the ideal cavity. Such backscattering suppression in the quantum regime with $N_{\mathrm{cw}}\ll 1$ [Fig.~\ref{Fig1}(c)] is quite different from that of classical light~\cite{Ang2017Fundamental,jaffe2022understanding,lee2023chiral,svela2020coherent,krenz2007controlling,Chen2021Synthetic,kim2019dynamic,orsel2023electrically,Jiang2024Coherent}.

Specifically, the behavior of $g^{(2)}_{\mathrm{cw}}(0)$ depends on both of $\chi$ and $\phi$, while $N_\mathrm{cw}$ is independent of the $\chi$ [Fig.~\ref{Fig2}]. By fixing $\phi=0.27\ \mu\mathrm{m}$, the classical revival of $N_\mathrm{cw}$ always exists with its maximum. In contrast, the quantum revival of SPB can only exist at the same position in the strong nonlinear regime ($\chi/\gamma>1$), but cannot exist for $\chi/\gamma<1$. In addition, by fixing $\phi=\{0.21,\ 0.33\}\ \mu\mathrm{m}$, $N_\mathrm{cw}$ cannot be totally revived, regardless of the strengths of the nonlinearity. However, SPB can emerge with a specific strength of the nonlinearity, $\chi/\gamma=0.5$.

%%%%%%%%%%%%%%%%%%%%%%%%%%%%%%%%%%%%%%%%%%%%%%%%%%%
\begin{figure}[b!]
    \includegraphics[width=0.95\textwidth]{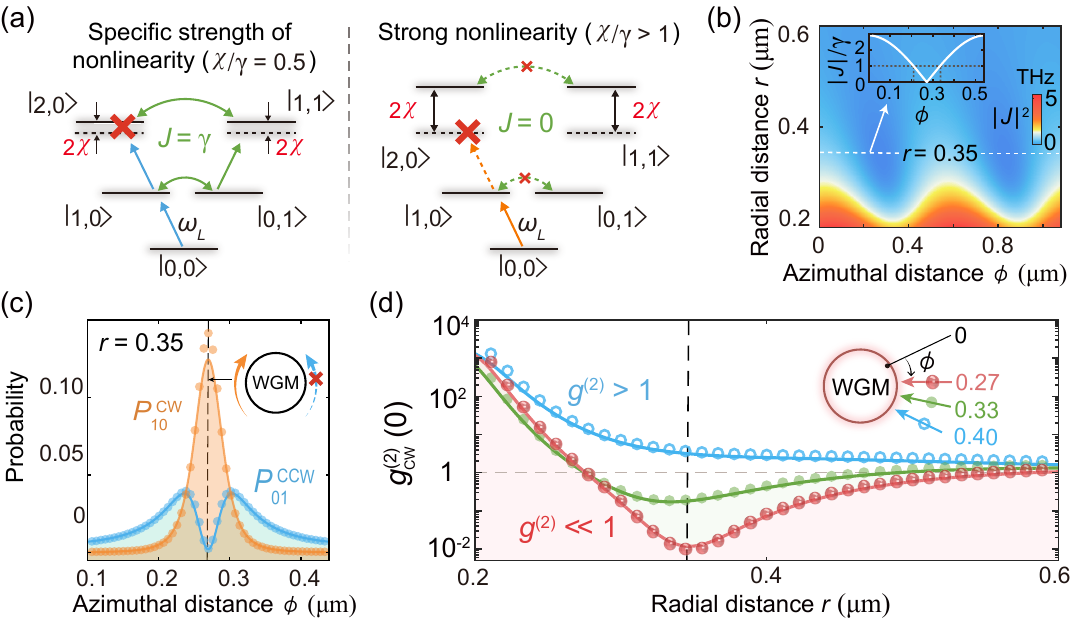}
    \caption{Physical mechanism. {(a)} Eigenenergy structures and transition paths for $\chi/\gamma=0.5$ and $\chi/\gamma>1$. {(b)} Total coupling strength $\vert J\vert^{2}$ versus $r$ and $\phi$. The inset shows $|J|/\gamma$ versus $\phi$ at $r=0.35$. (c) The $P_{01}^\mathrm{ccw}$ ($P_{10}^\mathrm{cw}$) is suppressed (revived) at $\phi=0.27\ \mu\mathrm{m}$. {(d)} The ${g}^{(2)}_\mathrm{cw}(0)$ versus $r$ with different $\phi$. In (c,d), the curves and markers correspond to the numerical and analytical results, respectively. The Kerr nonlinearity and other parameters are the same as those in Fig.~\ref{Fig1}.}
\label{Fig3}
\end{figure}
%%%%%%%%%%%%%%%%%%%%%%%%%%%%%%%%%%%%%%%%%%%%%%%%%%%
\subsection{Physical mechanism}
The underlying physics can be understood from the interplay of the resonator nonlinearity and the tip-induced optical coupling by analyzing the photon-number probabilities and the transitions between different quantum states [Fig.~\ref{Fig3}], which is different from that in previous studies in classical optics~\cite{Ang2017Fundamental,jaffe2022understanding,lee2023chiral,svela2020coherent,krenz2007controlling,Chen2021Synthetic,kim2019dynamic,orsel2023electrically,Jiang2024Coherent}, i.e., merely relies on the scatterer-induced destructive interference, and focuses on light amplitudes.

We study the photon-number probabilities via the quantum trajectory method~\cite{plenio1998quantumjump}. Our effective Hamiltonian is $\hat{H}_{\text{eff}}=\hat{H}_r-i\sum_{j=1,2}(\gamma/2)\hat{a}_{j}^{\dagger}\hat{a}_{j}$. For $\xi\ll\gamma$, by truncating the Hilbert space to $N=m+n=3$, the states are $|\psi(t)\rangle=\sum_{m=0}^{3}\sum_{n=0}^{m}C_{mn}|m,n\rangle$, where $C_{mn}$ are probability amplitudes corresponding to $|m,n\rangle$. The probability of finding $m$ photons in the CW mode and $n$ photons in the CCW mode is given by $P_{mn}=\vert C_{mn}\vert^2$, which can also be obtained from the steady-state solutions $\rho_\text{ss}$ of Eq.~(\ref{eq:ME}) via $P_{mn}=\langle m,n|\rho_\text{ss}|m,n\rangle$. An excellent agreement between our analytical results and numerical results is seen in Fig.~\ref{Fig3}. Note that the effect of quantum jumps is ignored (considered) in the semiclassical analytical (quantum master equation) approach~\cite{Minganti2019Quantum}.

For $\Delta=0$, the input light can be resonant with the transitions from the vacuum to $|2, 0\rangle$ in the weak nonlinear regime ($\chi/\gamma<1$). The corresponding probability amplitude can be obtained by solving the Schr\"odinger equation $i|\dot{\psi}(t)\rangle=\hat{H}_{\mathrm{eff}}|\psi(t)\rangle$:
\begin{align}\label{eq:C20}
C_{20}=\frac{2\sqrt{2}\xi^2(\Delta_1\Delta_2+4|J|^2\chi/\Delta_2)}{\eta_1\eta_2},
\end{align}
where $\Delta _{1}=2\Delta-i\gamma$, $\Delta _{2}=\Delta_{1}+2\chi$, $\eta_{1}=4\vert J\vert^2-\Delta _{1}^{2}$, and $\eta_{2}=4\vert J\vert^2-\Delta _{2}^{2}$. However, SPB can occur with $P_{20}=0$~\cite{liew2010single,bamba2011origin,Flayac2017Unconventional}, which can be understood from the destructive interference of two transition paths [Fig.~\ref{Fig3}(a)]: $|1,0\rangle\stackrel{\omega_L}{\longrightarrow}|2,0\rangle$ (blue), and $|1,0\rangle\stackrel{J}{\longrightarrow}|0,1\rangle\stackrel{\omega_L}{\longrightarrow}|1,1\rangle\stackrel{J}{\longrightarrow}|2,0\rangle$ (green). By setting $C_{20}=0$, the conditions of SPB are given by $\chi/\gamma=0.5$, and $|J|/\gamma=1$. For $J_0=1.8\gamma_1$, we have $r=0.35\ \mu\mathrm{m}$, and $\phi=\{0.21,\ 0.33\}\ \mu\mathrm{m}$ [the inset in Fig.~\ref{Fig3}(b)]. In contrast, $N_\mathrm{cw}$ cannot be totally revived at the same positions due to the nonzero coupling ($J\neq0$) between the CW and CCW modes.

The single-photon probabilities are given by:
\begin{align}\label{eq:P01}
P_{10}^{\mathrm{cw}}=4\xi^2\left\vert\frac{\Delta_1}{\eta_{1}}\right\vert^2,\quad P_{01}^{\mathrm{ccw}}=16\xi^2\left\vert\frac{J}{\eta_{1}}\right\vert^2,
\end{align}
where $P_{01}^{\mathrm{ccw}}$ tends to be zero, and $P_{10}^{\mathrm{ccw}}$ reaches its maximum for $J=0$ [Fig.~\ref{Fig3}(c)], i.e., $r={\ln({a_t/J_0})}/{2\alpha_\mathrm{t}}$, and $\phi={[(2l+1)\pi-\theta-\theta_tr]}/{2k_\mathrm{opt}}$ with integer $l$. For $l=\{0,1\}$, we have $\phi=\{0.27,\ 0.82\}\ \mu\mathrm{m}$ [Fig.~\ref{Fig3}(b)]. SPB emerges because the transition $|0,0\rangle\to|1,0\rangle$ is resonantly driven by the input light, but the transition $|1,0\rangle\to|2,0\rangle$ is detuned by $2\chi$, and the transitions between $|1,0\rangle$ and $|0,1\rangle$ are eliminated [Fig.~\ref{Fig3}(a), right panel]. Such effect can be understood from the interplay of the strong nonlinearity induced unequal eigenenergy spaces ($\chi/\gamma>1$), and the tip-induced vanishing of the coupling ($J=0$). In contrast, the classical revival of $N_\mathrm{cw}$ merely relies on the condition of $J=0$, regardless of $\chi$.

%\begin{align}\label{eq:critical}
%r=\frac{\ln({a_t/J_0})}{2\alpha_\mathrm{t}},\quad\phi=\frac{(2l+1)\pi-\theta-\theta_tr}{2k_\mathrm{opt}},
%\end{align}

The second-order quantum correlation is given by
\begin{align}\label{eq:GTA}
g^{(2)}_{\mathrm{cw}}(0)\simeq\frac{2P_{20}}{P_{10}^2}=\left\vert \frac{\eta _{1}\left( \Delta _{1}\Delta _{2}+4\left\vert J\right\vert ^{2}\chi /\Delta _{2}\right) }{\Delta _{1}^{2}\eta _{2}}\right\vert ^{2}.
\end{align}
With $J=0$, we obtain $g^{(2)}_{\mathrm{cw}}(0)\simeq[4(\chi/\gamma)^2+1]^{-1}<1$, for $\chi>\gamma$ and $\Delta=0$, which indicates SPB [Fig.~\ref{Fig3}(d)]. When the tip is away from the cavity ($r>0.6\ \mu\mathrm{m}$), the system behaves as a nonideal cavity without the tip and SPB cannot be observed. When the tip is close to the cavity ($0<r<0.2\ \mu\mathrm{m}$), increased $\gamma$ and $J$ enable the input light to be in resonance with higher photon-number states, resulting in photon bunching~\cite{Faraon2008}.

\subsection{Robust SPB for different backscattering strengths}
\begin{figure}[t!]
    \includegraphics[width=0.8\textwidth]{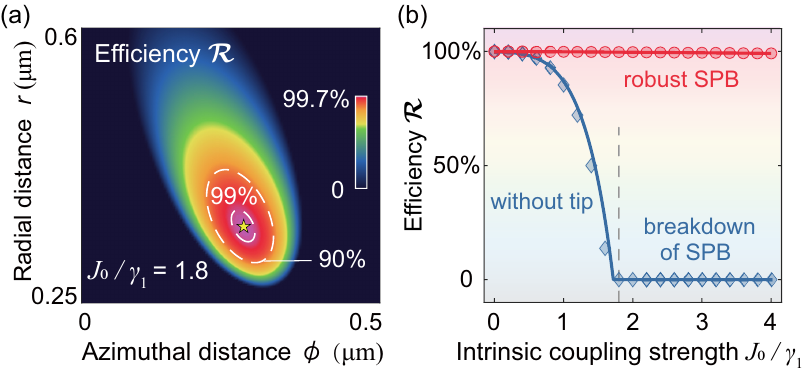}
    \caption{Robust SPB for different backscattering strengths. {(a)} SPB against backscattering loss efficiency $\mathcal{R}$ versus $r$ and $\phi$ for $J_{0}/\gamma_{1}=1.8$ and $\Delta=0$. {(b)} Efficiency $\mathcal{R}$ as a function of $J_{0}$ with (red) and without (blue) the nanotip, where the curves and markers show our numerical and analytical results, respectively. The Kerr nonlinearity and other parameters are the same as those in Fig.~\ref{Fig1}.}
\label{Fig4}
\end{figure}

To characterize the SPB against backscattering loss efficiency, we introduce a ratio by comparing the minimum of $g^{(2)}_{\mathrm{cw}}(0)$ in our device ($J_0\neq0$) with that in an ideal cavity ($J_0=0$) under the same optical nonlinearities and driving fields:
\begin{align}
\mathcal{R}=\max\,[0,\zeta], \quad\zeta\equiv\frac{1-\mathrm{min}[g^{(2)}_{\textrm{cw}}(0)(J_0\neq0)]}{1-\mathrm{min}[g^{(2)}_{\textrm{cw}}(0)(J_0=0)]}.
\end{align}
Here, the quantity of $1-\mathrm{min}[g^{(2)}_{\textrm{cw}}(0)]$ is the purity of the generated single photons, and $\mathcal{R}=100\%$ denotes perfect backscattering immunity, indicating that the single photons generated in our system with the intrinsic backscattering have the same purity as those in the ideal case. Figure~\ref{Fig4}(a) shows that the efficiency $\mathcal{R}$ can reach $99.7\%$ with $J_0/\gamma_1=1.8$ by adjusting the nanotip position at $r=0.35\ \mu\mathrm{m}$ and $\phi=0.27\ \mu\mathrm{m}$.
%%%%%%%%%%%%%%%%%%%%%%%%%%%%%%%%%%%%%%%%%%%%%%%%%%%
%%%%%%%%%%%%%%%%%%%%%%%%%%%%%%%%%%%%%%%%%%%%%%%%%%%
Furthermore, for the nonideal cavity without nanotip, such efficiency $\mathcal{R}$ gradually decreases with increasing $J_0$, and becomes 0 for $J_0/\gamma_1=1.8$ [Fig.~\ref{Fig4}(b)], i.e., SPB is suppressed by the backscattering. However, robust SPB can exist with different backscattering strenghths by introducing an additional tip with strong nonlinearities, which can be beneficial for protecting the generation or transmission of single photons, and improving the performance in realistic quantum devices.

\section{Conclusions and outlook}
We studied SPB against backscattering loss in a nonideal Kerr WGM cavity with a  nanotip. The efficiency of such effect is up to 99.7\% by tuning tip positions, which is robust with different backscattering strengths. More interestingly, we found that the behavior of this quantum effect is distinct from that of the classical mean-photon number with different strengths of the nonlinearity, due to the interplay of the resonator nonlinearity and the tip-induced optical coupling.

This underlying principle can be extended to other types of platforms, e.g., optical parametric amplifiers or cavity QED systems, for exploring squeezing or entanglement against backscattering loss, and for generating robust Schr\"odinger cat states. It is also expected to explore multi-photon bundles against backscattering loss~\cite{munoz2014emitters,Bin2020NPhonon,Bin2021Parity} or mutual blockade~\cite{Hamsen2018Strong} and robust microwave-optical photon pair~\cite{Painter2024Nonclassical} by studying higher-order correlations. Our work provides a novel perspective towards enhancing the performance of quantum devices, opening a new way to protect or engineer fragile quantum resources, and holding the potential for implementations in quantum technologies, such as robust single-photon routing~\cite{aoki2009efficient} in quantum communications or more robust quantum sensing~\cite{fleury2015invisible,yang2015invisible}.

\begin{acknowledgement}
H.J. is supported by the NSFC (Grant No. 11935006, 12421005), the Sci-Tech Innovation Program of Hunan Province (Grant No. 2020RC4047), the National Key R\&D Program (Grant No. 2024YFE0102400), and the Hunan Major Sci-Tech Program (Grant No. 2023ZJ1010). R.H. is supported by the RIKEN Special Postdoctoral Researchers (SPDR) program.  X.-W.X. is supported by the NSFC (Grants No.~12064010), and the science and technology innovation Program of Hunan Province (Grant No.~2022RC1203). A.M. is supported by the Polish National Science Centre (NCN) under the Maestro Grant no. DEC-2019/34/A/ST2/00081. F.N. is supported in part by: Nippon Telegraph and Telephone Corporation (NTT) Research, the Japan Science and Technology Agency (JST) [via the CREST Quantum Frontiers program Grant No. JPMJCR24I2, the Quantum Leap Flagship Program (Q-LEAP), and the Moonshot R\&D Grant No. JPMJMS2061], and the Office of Naval Research (ONR) Global (via Grant No. N62909-23-1-2074).

\end{acknowledgement}



\bibliography{references}

\end{document}